\documentclass[preprint]{aastex}
\usepackage{emulateapj5}

\slugcomment{To appear in The Astrophysical Journal Letters}
\shorttitle{The Hot PAGB Star ZNG 1 in M5}
\shortauthors{Dixon, Brown, \& Landsman}

\newcommand{\fig}[1]{Fig.~\ref{#1}}

\newcommand{\cfour}{\ion{C}{4}}
\newcommand{\cthree}{\ion{C}{3}$^\ast$}

\newcommand{\hetwo}{\ion{He}{2}}
\newcommand{\hone}{\ion{H}{1}}
\newcommand{\htwo}{H$_2$}
\newcommand{\nfive}{\ion{N}{5}}

\newcommand{\oone}{\ion{O}{1}}
\newcommand{\osix}{\ion{O}{6}}

\newcommand{\ebv}{$E($\bv)}

\newcommand{\kms}{km s$^{-1}$}
\newcommand{\logg}{$\log g$}

\newcommand{\msun}{$M_{\sun}$}
\newcommand{\rv}{$R_V$}
\newcommand{\specfit}{{\small SPECFIT}}
\newcommand{\teff}{$T_{\rm eff}$}
\newcommand{\vinf}{$v_{\infty}$}
\newcommand{\vlsr}{$V_{\rm LSR}$}
\newcommand{\vrot}{$v \sin i$}

\newcommand{\fuse}{{\it FUSE}}
\newcommand{\hst}{{\it HST}}

\newcommand{\iue}{{\it IUE}}

\received{2003 September 4}
\begin{document}

\title{The Rapidly Rotating, Hydrogen Deficient, Hot Post-Asymptotic
Giant Branch Star ZNG~1 in the Globular Cluster M5\footnotemark}

\footnotetext[1]{Based on observations made with the NASA-CNES-CSA {\it
Far Ultraviolet Spectroscopic Explorer.  FUSE}\/ is operated for NASA by
the Johns Hopkins University under NASA contract NAS5-32985.}

\author{W.\ Van Dyke Dixon}
\affil{The Johns Hopkins University, Department of Physics and Astronomy,
3400 North Charles Street, Baltimore, MD 21218} 
\email{wvd@pha.jhu.edu}

\medskip

\author{Thomas M. Brown}
\affil{Space Telescope Science Institute, 3700 San Martin Drive,
Baltimore, MD 21218}
\email{tbrown@stsci.edu}

\and

\author{Wayne B. Landsman}
\affil{Science Systems and Applications, Inc., Code 681,
NASA Goddard Space Flight Center, Greenbelt, MD 20771}
\email{landsman@mpb.gsfc.nasa.gov}

\begin{abstract}

We report observations of the hot post-asymptotic giant branch
star ZNG~1 in the globular cluster M5 (NGC~5904)
with the {\it Far Ultraviolet Spectroscopic Explorer (FUSE)}.\/ 
From the resulting spectrum, we derive 
an effective temperature \teff\ = $44300 \pm 300$ K,
a surface gravity \logg\ = $4.3 \pm 0.1$, 
a rotational velocity \vrot\ = $170 \pm 20$ \kms, and
a luminosity $\log L/L_\sun = 3.52 \pm 0.04$.  The atmosphere is
helium-rich (Y = 0.93), with enhanced carbon (2.6\% by mass), nitrogen
(0.51\%) and oxygen (0.37\%) abundances.   The spectrum shows evidence for a
wind with terminal velocity near 1000 \kms\ and an expanding shell of carbon-
and nitrogen-rich material around the star.
The abundance pattern of ZNG~1 is suggestive of the ``born-again'' scenario, 
whereby a star on the white-dwarf cooling curve undergoes a very late shell flash
and returns to the AGB, but the star's rapid rotation is more easily explained by
a previous interaction with a binary companion.

\end{abstract}


\section{INTRODUCTION}

Post-asymptotic giant branch (PAGB) stars represent a brief ($10^4$--$10^5$ yr)
phase of stellar evolution during which stars move rapidly
across the HR diagram at constant luminosity from the cool tip of the 
AGB to the beginning of the hot white-dwarf cooling curve.
As such, they are keys to both the past and the future:  their
atmospheric abundances provide important information about the mixing and mass-loss
processes at work in AGB stars, while their masses yield constraints on the
masses of white dwarfs currently forming in globular clusters.
To investigate these issues, we observed the PAGB
star ZNG~1 (Table~\ref{obs_parms}) 
in the globular cluster M5 (NGC~5904) with the {\it Far Ultraviolet
Spectroscopic Explorer (FUSE).}

First identified as a UV-bright star by
\citet*{ZNG:72} on the basis of its extreme $U-V$ color, ZNG~1 was observed
with \iue\/ by \citet{Bohlin:83}, who suggested that an apparent \ion{N}{4}
$\lambda 1487$ emission feature is the signature of a planetary nebula (PN). 
Combining this spectrum with other archival \iue\/ data, \citet{deBoer:85}
argued that the apparent \ion{N}{4} feature is spurious and pointed out the
P-Cygni shape of the strong \ion{N}{5} $\lambda \lambda 1238, 1242$ doublet.
To test the PN hypothesis, \citet{Napiwotzki:97} used the {\it Hubble
Space Telescope} to obtain a WFPC2 H$\alpha$ image and GHRS ultraviolet
spectra of the star.  The H$\alpha$ image shows a bright G-type star only 0\farcs5
from ZNG~1, but no evidence for extended emission around the star.  The
GHRS spectra clearly show P-Cygni profiles in the \nfive\ doublet and
discrete absorption components (DACs), blueshifted by $\sim$ 900 \kms,
in both \nfive\ and \cfour\ $\lambda \lambda 1548, 1550$.  No nebular emission
is present.

\section{OBSERVATIONS AND DATA REDUCTION}

\fuse\/ consists of four separate optical systems.  Two employ LiF
optical coatings and are sensitive to wavelengths from 990 to 1187 \AA,
while the other two use SiC coatings, which provide reflectivity to
wavelengths as short as 905 \AA.  The four channels overlap between 990
and 1070 \AA.  The \fuse\/ flux calibration, based on theoretical models of
white-dwarf stellar atmospheres, is believed accurate to about 10\%.
For a complete description of \fuse, see \citet{Moos:00} and \citet{Sahnow:00}.

The \fuse\/ spectrum of ZNG~1 in M5 (data set A1080303) was
obtained in 4 separate exposures on 2000 July 15.  The total
integration time was 3681 s, all of it during orbital night.  All
observations were made through the 30\arcsec\ $\times$ 30\arcsec\ (LWRS)
aperture.  Archival \hst/WFPC2 FUV images of the cluster core confirm that
no other UV-bright stars fell within the spectrograph aperture.
The data were reduced using version 2.3 of the CalFUSE calibration software pipeline.
For each \fuse\/ channel, the spectra from individual exposures 
were cross-correlated, shifted to
a common wavelength scale, weighted by their exposure time, and
averaged.  The spectrum from each channel was then cross-correlated
with a synthetic molecular-hydrogen spectrum and shifted to match.  The
wavelength scale defined in this way is at rest with respect to the
interstellar medium (ISM), which we take as a proxy for the local
standard of rest.  Finally, the spectra were binned by four detector
pixels, or about 0.025~\AA, approximately half of the instrument resolution.

\section{SPECTRAL ANALYSIS}
\label{analysis}

\subsection{Synthetic Spectra and Model Fitting}
\label{synthetic}

Model atmospheres are computed using the program TLUSTY \citep{Hubeny:Lanz:95},
which employs a sophisticated treatment of non-LTE (local thermodynamic equilibrium)
line blanketing to account for the opacity sources important in stellar atmospheres.
In our models, the line and continuum transitions of
H, He, C, and N (a total of 10 ions and 220 energy levels) are treated
in non-LTE with detailed photoionization cross sections that include resonances;
other species, assumed to have scaled-solar abundances with [Fe/H] equal to the 
cluster mean (Table \ref{obs_parms}), are treated in LTE.
All models assume a microturbulent velocity of 2 \kms.
Synthetic spectra are generated using the program SYNSPEC and rotationally
broadened using the program ROTIN3, which assumes constant limb darkening.
The application of these codes to another sdO star, BD~$+75$\arcdeg 325, is described
in \citet*{Lanz:97}.

Model spectra are fit to the data using the nonlinear curve-fitting
program \specfit\ \citep{Kriss:94}, which 
performs a $\chi^2$ minimization of the model parameters.
Reddening is modeled with
a \citet*{CCM:89} extinction curve, extrapolated to the Lyman limit,
assuming \ebv\ = 0.03 (Table \ref{obs_parms}) and \rv\ = 3.1.  Error bars for
a particular parameter are derived by fixing that parameter at the
best-fit value, then raising it, while allowing the other model
parameters to vary freely, until $\chi^2$ rises by 1, which corresponds
to a 1 $\sigma$ deviation for a single interesting parameter
\citep{Avni:76}.  The error bars quoted in Table \ref{stellar_parms} 
are purely statistical and do not include systematic errors.  

\subsection{Atmospheric Parameters}

\begin{figure}
\epsscale{0.70}
\plotone{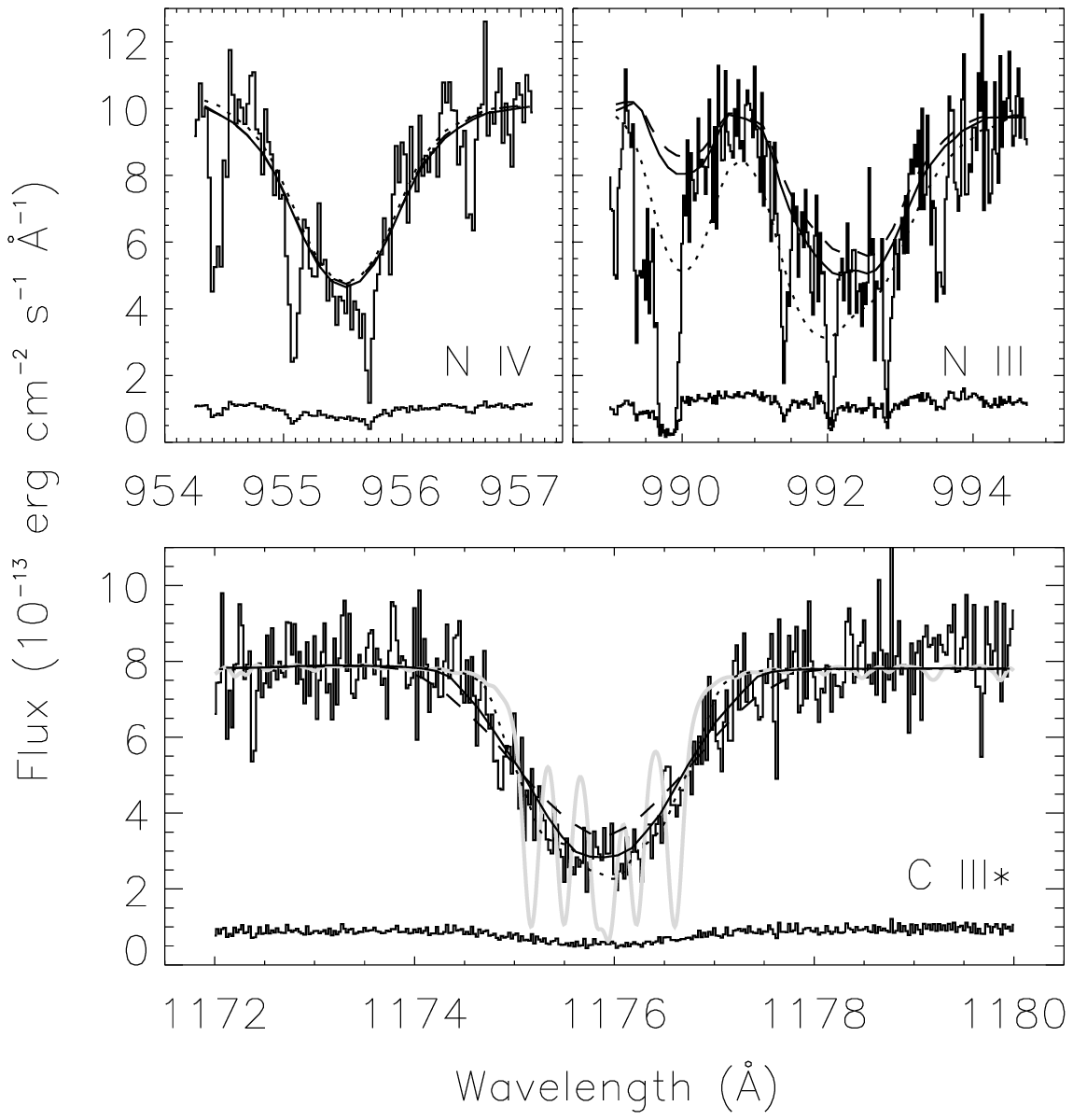}
\caption{
Nitrogen and carbon features in the \fuse\/ spectrum of ZNG~1 in M5.
Data are binned to 0.025 \AA\ and plotted as histograms.
Error bars and synthetic spectra are overplotted.
{\it Top panel:} The \ion{N}{4} $\lambda 955$, \ion{N}{3} $\lambda 990$,
and \ion{N}{3} $\lambda 992$ lines overplotted by models with
\teff\ = 40 (dotted line), 45 (solid line), and 50 kK (dashed line).
The narrow absorption features are due to interstellar \htwo\ and \oone.
{\it Bottom panel:}  The \ion{C}{3}* $\lambda 1175$ multiplet overplotted
by models with rotational velocity $v \sin i = 10$ (gray
line), 100 (dotted line), 200 (solid line), and 300 (dashed line) \kms.
\label{carbon}}
\end{figure}

The most prominant feature in the \fuse\/ spectrum of ZNG~1 is the
\cthree\ multiplet at 1175 \AA.  The half dozen
individual lines of this feature are generally well resolved in
\fuse\/ spectra, but in ZNG~1 are blended into a broad absorption
trough.  As shown in \fig{carbon}, this feature is well
reproduced by a rotational velocity \vrot\ = 170 \kms,
a carbon mass fraction of 3\%, an effective temperature
\teff\ = 45 kK, and a radial velocity \vlsr\ = +60 \kms, a value
consistent with the cluster velocity (Table \ref{obs_parms}).

\begin{figure}
\epsscale{0.70}
\plotone{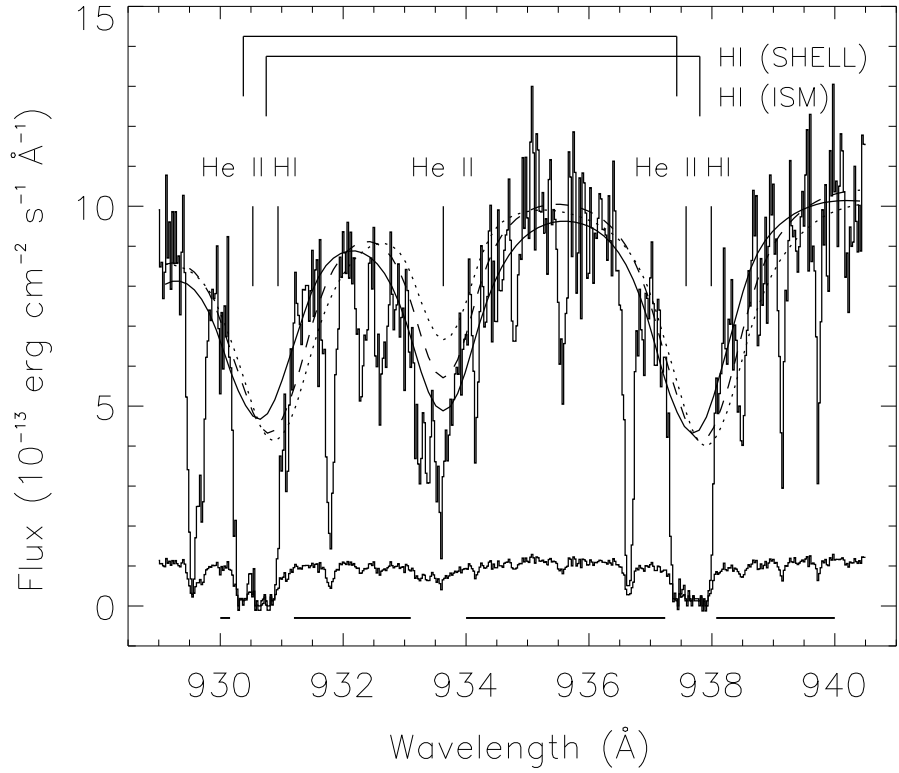}
\caption{
Hydrogen and helium lines in the \fuse\/ spectrum of ZNG~1 in M5.  Data are
taken from the two SiC detector segments, binned to 0.025 \AA, and
plotted as a histogram.  Error bars and synthetic spectra are
overplotted.  The thick, solid lines below the spectrum mark the spectral 
regions included in the model fits.
The models have \teff\ = 45 kK, $\log g = 4.5$, and helium
fractions of 10\% (dotted line), 40\% (dashed line), and 90\% (solid
line) by number.  As the helium fraction increases, the helium-hydrogen blends
at 931 and 938 \AA\ shift to the blue and the
\hetwo\ $\lambda 934$ line becomes stronger.
Above the spectrum are marked two pairs of foreground \hone\ absorption lines;
the bluer pair may represent a shell of material expanding about the star.
\label{helium}}
\end{figure}

We determine the star's surface gravity and helium abundance 
from simultaneous fits to its helium and hydrogen lines.  
As shown in \fig{helium}, every second \hetwo\ Balmer line is paired with
an \hone\ Lyman line, but lies $\sim 0.5$ \AA\ to the blue.  
The lines are unresolved because of the star's high rotational velocity,
but our models show that these H-He blends shift to the
blue as the helium abundance rises and the hydrogen abundance falls.
(Note that the stellar features are redshifted relative to the
ISM.)  Though the cores of all three features are contaminated---the
H-He blends by interstellar \hone\ and the \hetwo\ $\lambda 933$ 
line by \ion{S}{6} from the stellar wind---their wings can constrain both
the stellar surface gravity and the helium abundance.

We model the region between 930 and 940 \AA, excluding the cores
of all three stellar features (\fig{helium}).  (Model fits to the ISM
lines confirm that all regions of significant foreground
\hone\ absorption are excluded.)  The remaining interstellar lines
are fit with simple Gaussians.  
We generate a grid of models with
surface gravities \logg\ = 4.0, 4.5, and 5.0 and helium fractions from 10 to 90\%
(in steps of 10\%) by number.  Within \specfit, we linearly
interpolate between models to produce a fine grid with steps of 0.1 dex
in \logg. 
Assuming \teff\ = 45 kK and \vlsr\ = +60 \kms, we find a surface gravity
of $\log g = 4.3 \pm 0.1$ and a helium fraction of $85 \pm 5$\% by number.
Repeating this test with \teff\ = 46 kK models
yields best-fit values of $\log g = 4.4 \pm 0.1$ and a helium fraction
of $85 \pm 5$\%, so we need not increase our error bars to account
for small uncertainties in the effective temperature.
Given the low hydrogen abundance of this star, we quote all abundances
as mass fractions; that of helium is $93 \pm 2$\% (Table~\ref{stellar_parms}).

Adopting these stellar parameters, we fit model spectra to
four photospheric nitrogen features,
\ion{N}{4} $\lambda 955$,
\ion{N}{3} $\lambda 980$,
\ion{N}{3} $\lambda 990$ (red wing), and
\ion{N}{3} $\lambda 992$,
and derive an effective temperature of \teff\ = $44300 \pm 300$ K and
a nitrogen mass fraction of $0.51 \pm 0.05$\%.
Fits to three of these lines are presented in \fig{carbon}.
The giants in M5 show a wide range of nitrogen abundances, 
the highest being [N/Fe] = +1.38 \citep*{Cohen:02},
but the nitrogen abundance of ZNG~1 is 4 times greater.  
Returning to the \cthree\ multiplet, we obtain a rotational velocity
$v \sin i = 170 \pm 20$ \kms\ and
a carbon mass fraction of $2.58 \pm 0.20$\%, more than 
160 times greater than the most carbon-rich
cluster giant studied by \citet{Cohen:02}.

The star's \fuse\ spectrum shows no strong oxygen features, so we fit the wavelength
region between 1090 and 1104 \AA, where the continuum is fairly flat and
where numerous weak \ion{O}{3} and \ion{O}{4} lines combine to form two
broad absorption troughs centered at 1093 and 1099 \AA.  We derive an
oxygen mass fraction of $0.37 \pm 0.32$\%.    While the oxygen abundance
is the least well-determined of the CNO abundances in ZNG~1, it is certainly  larger
than the maximum value observed in M5 giants, where [O/Fe] ranges from $-0.55$ to
$+0.47$ \citep{Ivans:01}.
The star also appears to be enhanced in S, Si, and P; we will determine their
abundances in a future paper.  

\subsection{Wind and Shell}

The \fuse\/ spectrum of ZNG~1 shows broad P-Cygni profiles and blueshifted
discrete absorption components (DACs)
in both components of the \osix\ doublet. 
The \osix\ profiles are well fit by models with a terminal velocity of \vinf\ = 1000 \kms.
The DACs, which are generally attributed to
density enhancements in the stellar wind,
are blueshifted by 960 \kms\ relative to the star, consistent with the
DACs seen by GHRS \citep{Napiwotzki:97}.
The spectrum also shows narrow absorption features of \hone,
\ion{C}{3} $\lambda 977.02$, \ion{C}{2} $\lambda 1036.33$, and
\ion{N}{2} $\lambda 1083.99$ blueshifted by
$\sim$ 120 \kms\ relative to the ISM and
$\sim$ 180 \kms\ relative to the star
(two blue-shifted \hone\ features are marked in \fig{helium}; 
see also Fig.\ 1 of \citealt*{Dixon:03}).
These features may represent a shell of material expanding about the star. 
A complete analysis of the star's wind and shell features is forthcoming.

\subsection{Stellar Parameters}
\label{stellarparms}

Using the cluster distance and the ratio of observed to synthetic
spectra \citep*{Dixon:94}, we derive a stellar luminosity 
$\log L/L_\sun = 3.52 \pm 0.04$. The star lies on the
\citet{Schoenberner:83} PAGB evolutionary tracks and
should have a mass $M_* \sim 0.565$. The mass calculated from the parameters of
our best-fit model,  $M_* = 0.69 \pm 0.17 \; M_\sun$, 
is consistent with the Sch\"{o}nberner prediction.

\section{DISCUSSION}

Hot PAGB stars in globular clusters differ from their field counterparts in that they
rarely show evidence for third dredge-up (by which the products of helium
burning are brought to the surface) and are seldom accompanied by PNe
\citep{Jacoby:97}.  These characteristics can be understood qualitatively
as a result of the expected low core mass ($< 0.55  M_\sun$) of
globular cluster AGB stars. While the minimum core mass for third dredge-up is
still a subject of active research \citep*{Karakas:02},
nearly all current AGB models agree that third dredge-up should not occur in stars with
initial mass $\lesssim 0.8 \; M_\sun$.     The rarity of PNe can be explained by the lower
luminosities and longer crossing times of PAGB stars with low core masses;
the nebula disperses before the central star becomes hot
enough to ionize it \citep{Jacoby:97}.   \citet*{Alves:00} therefore suggest
that binary evolution is necessary for the formation of PNe in
globular clusters.      Interestingly, the best-studied PN in a globular cluster,
K648 in M15, shows clear evidence for third dredge-up in its central star
\citep*{Rauch:02}, but not the strong helium and nitrogen
enhancements seen in ZNG~1.

One way to produce a hydrogen-deficient star with enhanced CNO abundances
is for the star to experience a very late shell flash while descending the
white dwarf cooling curve and subsequently return to the AGB
\citep{Herwig:99}.     This ``born-again'' scenario might explain the absence
of a PN surrounding ZNG~1, because the star is so old (since it is on a
second crossing), that any nebula has long since dispersed.     For the
born-again scenario to work for ZNG~1, the hydrogen must not be completely burned,
and the carbon enhancement must not be too high.  \citet{Lawlor:03} provide a
born-again model of an initial 1 $M_{\sun}$ star with  Z = 0.001 that predicts stellar
parameters similar to those of ZNG~1 (see their Table 3),
although the CNO abundances are about a factor of three larger.     The
difficulty with the born-again scenario (or any single-star evolution scenario)
is providing an explanation for the observed rapid rotation.  
One possibility: to explain the observed rotational velocity of up to 40 \kms\ in some globular horizontal-branch stars \citep{behr:03},  \citet{sills:00} postulate that
red giants may have a rapidly rotating core.  If so, then
perhaps its angular momentum can be transferred to the surface during the 
convective mixing in the final shell flash.  

Might the properties of ZNG~1 be explained through interaction with another star?
First, we note that its very blue \bv\ color indicates that any
current companion must be quite faint.
Given the proximity of the star to the projected cluster center (4\farcs2)
and the moderately high stellar interaction rate in M5 \citep{Pooley:03},
one possiblity is a stellar collision, which is
thought to be the primary production mechanism for the blue stragglers found at
the centers of globular clusters.  Collisionally-created blue stragglers are
expected to be born with a high angular momentum.  \citet*{Shara:97}
measure a rotation rate of $155 \pm 55$ \kms\ and estimate a mass of
$1.7 \pm 0.4 \; M_{\sun}$ for a blue straggler in 47 Tuc.  A star of this mass 
would likely experience third dredge-up on the AGB, but would
also be expected to eject a PN, which is not seen for ZNG~1.  Furthermore,
it is unclear how much of a blue straggler's angular momentum
would survive its post-main sequence evolution.

A second scenario, in which a low-mass companion spirals into the AGB progenitor,
has been proposed to explain K648 in M15 by \citet{Alves:00}
and discussed in general by \citet{DeMarco:02}.
The increased AGB mass due to the ingestion of the companion might be
sufficient to allow third dredge-up to occur, but additional mechanisms such as
shear or rotationally-induced mixing are required to create a hydrogen-deficient
star \citep{DeMarco:02}.  A merger process is thought to be the origin of the
high rotational velocity of FK Comae stars ($v \sin i = 162.5 \pm 3.5$ \kms\ for
the prototype FK Comae; \citealt{Huenemoerder:93}).
More detailed analysis is required to assess the viability of a merger scenario
for ZNG~1; whether the merger remnant should eject a PN is of particular interest.

\citet{Pauldrach:88} compute terminal velocities and mass-loss rates
for the winds of (Pop.\ I) central stars of PNe.    Assuming an evolutionary mass of 
0.565 $M_{\sun}$ (\S \ref{stellarparms}) and  temperature of \teff\ = 45 K, their 
models predict mass-loss rates of a few times $10^{-9}$ $M_{\sun}$ yr$^{-1}$.
Given this modest wind,  it appears unlikely that ZNG~1 will dissipate
significant angular momentum in the brief time ($\sim 15,000$ years, according
to the 0.565 \msun\ track of \citealt{Schoenberner:83})
prior to its arrival on the white-dwarf cooling curve.
Previous studies have yet to find white dwarfs with such
rapid rotation (see review by \citealt{Kawaler:03}) but the example of ZNG~1
suggests that further searches may be fruitful. 



\acknowledgments

This research has made use of NASA's Astrophysics Data System
Bibliographic Services and the Catalogue Service of the CDS,
Strasbourg, France.  We thank I. Hubeny for supplying his spectral
synthesis codes 
and A. Sweigart for illuminating discussions.  We
acknowledge the outstanding efforts of the \fuse\/ P.I.\ team to make
this mission successful.  This work is supported by NASA grant NAG
5-10405.


\clearpage

\clearpage 

\begin{deluxetable}{lcc}
\tablewidth{0pt}
\tablecaption{Observational Parameters\label{obs_parms}}
\tablehead{
\colhead{Parameter} & 
\colhead{Value} &
\colhead{Reference} 
}
\startdata
Spectral type           & sdO           & 1 \\
$V$                     & 14.54          & 2 \\
\bv                     & $-0.32$        & 2 \\
$E$(\bv)                & 0.03          & 3 \\
Distance (kpc)          & 7.5           & 3 \\
{[Fe/H]}$_{\rm cluster}$ & $-1.27$      & 3 \\
$V_{{\rm {LSR, cluster}}}$ & 65.7     & 3 
\enddata

\tablerefs{(1) \citealt{Napiwotzki:97}, (2) \citealt{Piotto:02}, (3) \citealt{Harris:96} (on-line version dated February 2003).}

\end{deluxetable}

\clearpage

\begin{deluxetable}{lc}
\tablewidth{0pt}
\tablecaption{Derived Stellar Parameters\label{stellar_parms}}
\tablehead{
\colhead{Parameter} & \colhead{Stellar Value}
}
\startdata
\teff\		&	$44300 \pm 300$ K \\
\logg\		&	$4.3 \pm 0.1$ \\
$v \sin i$	&	$170 \pm 20$ \kms \\
\vlsr\		&	$+60 \pm 4$ \kms \\
$\log L/L_{\sun}$ &	$3.52 \pm 0.04$ \\
$M/M_{\sun}$ 	&	$0.69 \pm 0.17$ \\
Y         &	$0.93 \pm 0.02$ \\
C abundance	&	$2.58 \pm 0.20$\% \\
N abundance	&	$0.51 \pm 0.05$\% \\
O abundance     &       $0.37 \pm 0.32$\%
\enddata

\tablecomments{Abundances are quoted as mass fractions.}

\end{deluxetable}

\end{document}